\documentclass{webofc}
\usepackage[varg]{txfonts}   
\usepackage{lipsum}
\usepackage{xcolor}
\usepackage{indentfirst}
\usepackage{mdframed}
\usepackage{color,soul}
\usepackage{array}
\usepackage{wrapfig}
\usepackage[utf8]{inputenc}
\usepackage{aasmacros}
\begin{document}
\title{Measuring the CMB primordial B-modes with Bolometric Interferometry}
%
%
\subtitle{Status and future prospects of the QUBIC experiment}

\author{\lastname{A.~Mennella}\inst{1,2}\fnsep\thanks{\email{aniello.mennella@fisica.unimi.it}} \and
\lastname{P.~Ade}\inst{3}\and
\lastname{A.~Almela}\inst{4}\and
\lastname{G.~Amico}\inst{5}\and
\lastname{L.H.~Arnaldi}\inst{6}\and
\lastname{J.~Aumont}\inst{7}\and
\lastname{S.~Banfi}\inst{8,9}\and
\lastname{E.S.~Battistelli}\inst{5,10}\and
\lastname{B.~Bélier}\inst{11}\and
\lastname{L.~Bergé}\inst{12}
% ...
}

\institute{Università degli studi di Milano, Milano, Italy 
\and
	Istituto Nazionale di Fisica Nucleare (INFN) sezione di Milano, Milano, Italy 
\and
	Cardiff University, Cardiff CF10 3AT, United Kingdom 
\and
	Instituto de Tecnolog\a'{i}as en Detecci\a'{o}n y Astropart\a'{i}culas  (CNEA, CONICET, UNSAM), Argentina 
\and
	Universit\a`{a} di Roma - La Sapienza, Italy 
\and
	Centro At\a'{o}mico Bariloche and Instituto Balseiro (CNEA), Argentina 
\and
	Institut de Recherche en Astrophysique et Plan\a'{e}tologie, Toulouse (CNRS-INSU), France 
\and
	Universit\a`{a} di Milano - Bicocca, Milano, Italy 
\and
    INFN Milano-Bicocca, Milano, Italy 
\and
	INFN sezione di Roma, Roma, Italy 
\and
	Centre de Nanosciences et de Nanotechnologies, Orsay, France 
\and
	Centre de Spectrométrie Nucléaire et de Spectrométrie de Masse, Orsay, France 
% ...
}

\abstract{
The Q\&U Bolometric Interferometer for Cosmology (QUBIC) is the first bolometric interferometer designed to measure the primordial $B$-mode polarization of the Cosmic Microwave Background (CMB). Bolometric interferometry is a novel technique that combines the sensitivity of bolometric detectors with the control of systematic effects that is typical of interferometry, both key features in the quest for the faint signal of the primordial $B$-modes. A unique feature is the so-called ``spectral imaging'', i.e., the ability to recover the sky signal in several sub-bands within the physical band during data analysis. This feature provides an in-band spectral resolution of $\Delta\nu/\nu\sim 0.04$ that is unattainable by a traditional imager. This is a key tool for controlling the Galactic foregrounds contamination. In this paper, we describe the principles of bolometric interferometry, the current status of the QUBIC experiment and future prospects.}

\maketitle
%


\section{Introduction}
\label{sec-intro}

    Bolometric interferometry is a measurement technique implemented by the Q\&U Bolometric Interferometer for Cosmology (QUBIC), designed to measure the polarization $B$-modes of the cosmic microwave background (CMB). This signal, if detected, would provide undisputed evidence of a background of gravitational waves in the primeval universe and a clear signature of inflation.

The control of instrumental systematic effects and the removal of astrophysical foregrounds are the main challenge in the quest for CMB $B$-modes. Their sheer faintness imply a correspondingly high level of precision in the removal of polarized foregrounds, which are at least one order of magnitude larger than the cosmological signal.

After the claimed CMB $B$-modes detection by BICEP2 in 2014 turned out to be thermal dust emission \cite{bicep2-2014,Ade2015}, there has been an increased awareness in the importance of understanding and controlling polarized Galactic foregrounds. Analysis of WMAP, Planck and S-PASS data \cite{Krachmalnicoff2016} has revealed that no sky region can be considered clean with the target tensor-to-scalar ratio, $r$, sensitivity of 10$^{-3}$ in the next generation CMB experiments. Moreover, recent analysis of Planck data \cite{Pelgrims_2021} has shown that the spectrum of dust emission is more complex than that of a modified black body, usually assumed in parametric component separation algorithms.

Bolometric interferometry, with its spectral imaging capability \cite{2020.QUBIC.PAPER2}, provides a unique approach to tackle foreground issues, allowing one to detect complex spectral behavior in the dust emission that would be undetectable by imagers integrating the signal over bandwidths of several GHz. QUBIC is the first CMB instrument based of bolometric interferometry \cite{2020.QUBIC.PAPER1}, and a technological demonstrator has been recently deployed at Alto Chorrillo, in the Argentinean Andes, at $\sim$5000\,m a.s.l. In this paper we describe the current status of QUBIC and show the scientific potential of bolometric interferometry when compared  with an direct imaging experiment.

\section{Bolometric interferometry and QUBIC}
\label{sec-instrument}

\begin{figure}[h!]
        \begin{center}
            \includegraphics[width=\textwidth]{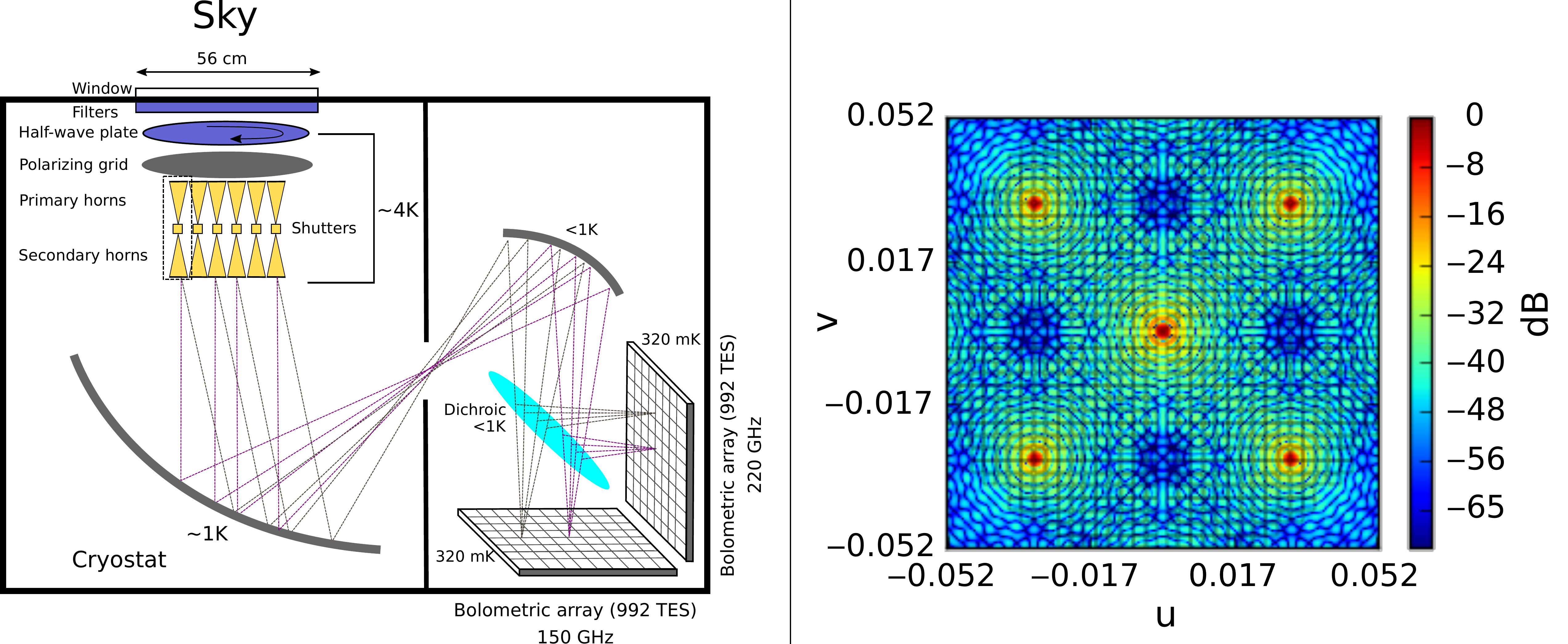}
        \end{center}
        \caption{\label{fig-instrument}\textit{Left panel}: schematic of the QUBIC instrument, also displaying the principle of bolometric interferometry. \textit{Right panel}: the pattern formed on the focal plane when the instrument observes a monochromatic point source along the optical axis.}
    \end{figure}

    The left panel of Fig.~\ref{fig-instrument} shows a schematic of QUBIC, which also highlights the principle of bolometric interferometry. The sky signal enters a cryostat through a window and then propagates through a 4\,K stage with thermal filters, a rotating half-wave plate and a polarizing grid that selects one of the two polarization directions. A set of 400+400 corrugated, back-to-back feedhorns interspersed with mechanical shutters transmits the field onto a 1\,K dual reflector optical combiner. A dichroic filter selects two 40\,GHz frequency bands centered at 150 and 220\,GHz that are detected by two arrays of 992 transition-edge sensors (TES), each cooled to 300\,mK.

On each of the two focal planes we obtain a pattern resulting from the additive interference of the fields transmitted by the feedhorns rather than the direct image of the sky. The right panel of Fig.~\ref{fig-instrument} shows the pattern obtained when the instrument observes a monochromatic point source located along the optical axis. This pattern, called \textit{synthetic beam}, essentially convolves the sky image, and displays a main peak with several secondary peaks arising from the constructive interference of the fields transmitted by the apertures.

\begin{figure}[h!]
        \begin{center}
            \includegraphics[width=13cm]{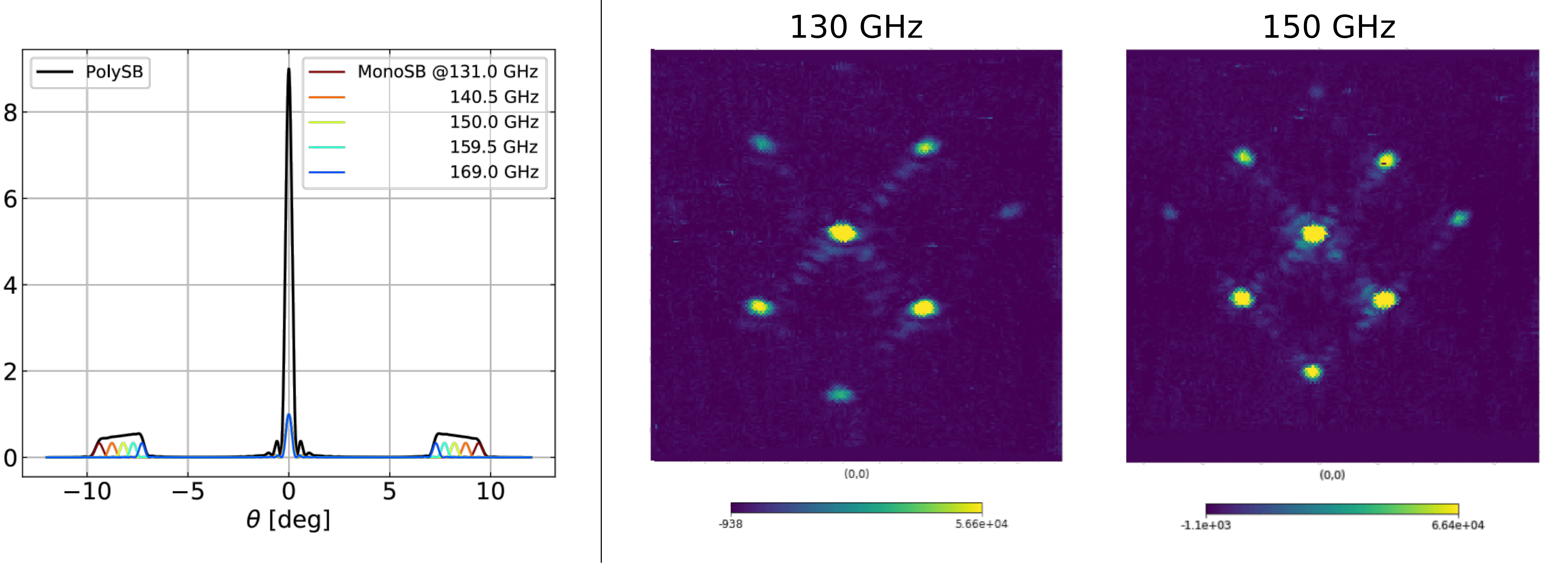}
        \end{center}
        \caption{\label{fig-synthetic_beam}\textit{Left panel}: simulation of a polychromatic synthetic beam (black curve) compared to a series of monochromatic beams at various frequencies (colored curves). The polychromatic beam has elongated secondary peaks that can be modelled as a sum of monochromatic peaks. This allows to resolve sub-frequencies. \textit{Right panel}: synthetic beams measured at 130 and 150\,GHz during laboratory calibration.}
    \end{figure}

When the source is not monochromatic the synthetic beam side peaks broaden, as shown in the cut displayed in the left panel of Fig.~\ref{fig-synthetic_beam}. This dependence of the secondary peaks from the signal frequency spectrum confers bolometric interferometry the ability to resolve sub-frequencies in the physical band and to work as a spectrometer as well as an imager. The number of sub-bands is selected in software during data analysis, so that the same data can be analyzed with various spectral combinations. The right panel of Fig.~\ref{fig-synthetic_beam} shows synthetic beams at 130 and 150\,GHz measured during laboratory calibration ~\cite{2020.QUBIC.PAPER3}. One can observe the varying position of the secondary peaks based on the signal frequency.

The presence of mechanical shutters in the middle of each horn pair allows to conduct \textit{self-calibration} \cite{Bigot-Sazy2013}. This procedure, typical of interferometers, allows one to calibrate a large number of instrument parameters and achieve strict control of systematic effects. The interested reader can find more details about the instrument and its laboratory calibration in \cite{2020.QUBIC.PAPER1,2020.QUBIC.PAPER2,2020.QUBIC.PAPER3,2020.QUBIC.PAPER4,2020.QUBIC.PAPER5,2020.QUBIC.PAPER6,2020.QUBIC.PAPER7,2020.QUBIC.PAPER8}.

\section{QUBIC expected science}
\label{sec-science}

\begin{figure}[h!]
    \begin{center}
     \includegraphics[width=\textwidth]{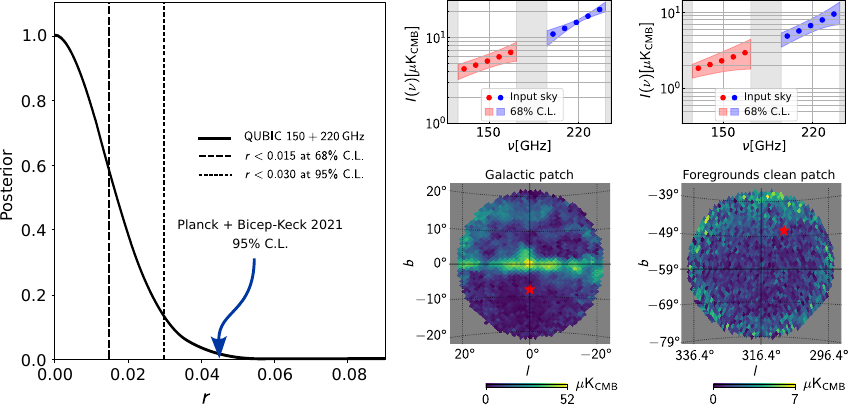}
     \caption{\label{fig-posterior+dust}\textit{Left}: posterior likelihood on the tensor-to-scalar ratio, $r$, obtained by QUBIC after three years of integration. \textit{Right}: dust emission SED reconstructed by QUBIC within its two physical bands in two pixels (red stars) sampled from two different sky patches. The points are the input sky intensities, the color bands represent the 68\% confidence level for the reconstructed SED. Figure adapted from \cite{2020.QUBIC.PAPER1}.}
    \end{center}
\end{figure}

    In Fig.~\ref{fig-posterior+dust} (adapted from \cite{2020.QUBIC.PAPER1}) we show a summary of the expected impact on science of QUBIC in terms of reconstruction of the tensor-to-scalar ratio, $r$, and of dust spectral characterization after three years of observations. 
The left panel displays the posterior likelihood on $r$ (assuming no foregrounds) obtained using both frequency bands, with the blue arrow indicating the latest constraint from Tristram et al. \cite{tristram2021}. 
The right panel shows the capability of QUBIC in detecting the spectrum of dust emission in polarization by resolving sub-frequencies within the physical band. The two columns represent the dust spectral energy density (SED) in two pixels sampled from two different sky patches: one close to the Galactic plane center (left column) and the other in a foregrounds-clean region (right column).

In presence of a potential detection of $r$, the ability to measure the dust emission SED with $\sim$\,GHz spectral resolution is key to assess the presence of foreground residuals if their spectrum is more complex than that assumed in component separation. To assess this we performed end-to-end simulations in which we compared an instrument with the sensitivity of CMB-S4 with a similar instrument with the ability to resolve sub-frequencies in the physical band.

\begin{figure}[h!]
    \begin{center}
     \includegraphics[width=12.8cm]{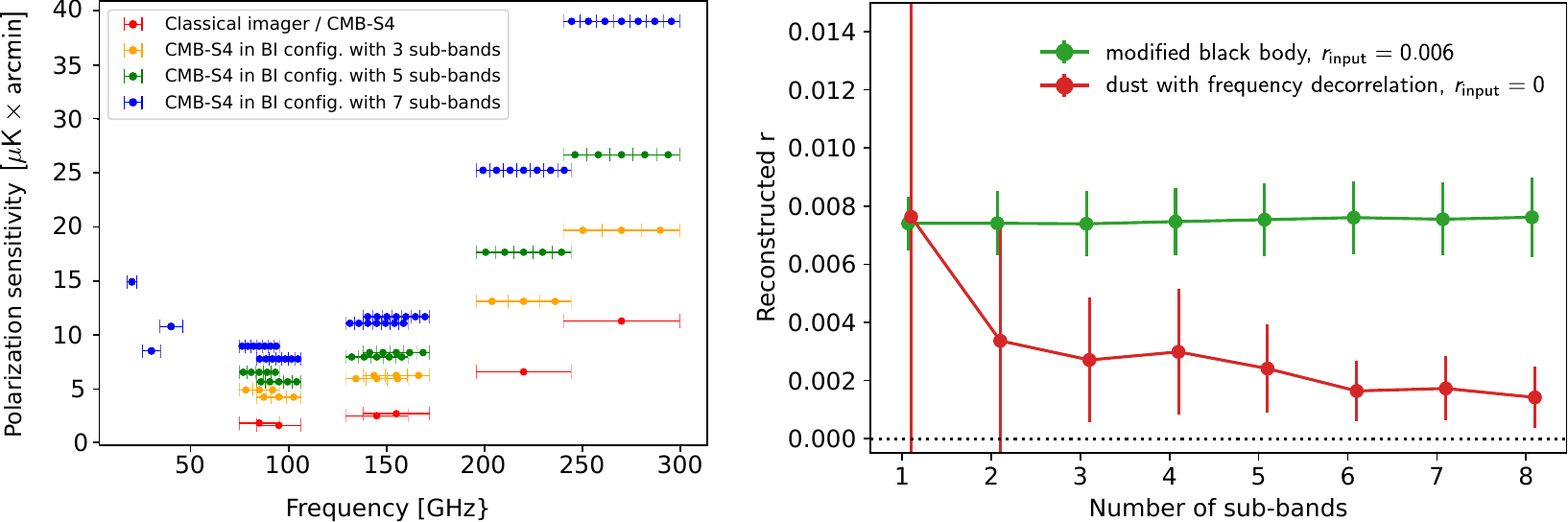}
     \caption{\label{fig-rbias}\textit{Left}: bandwidths and sensitivities of the tested configurations. \textit{Right}: value of the reconstructed $r$ as a function of $n_\mathrm{sub}$. Green line: dust with a modified black-body spectrum (PySM model \texttt{d1}). Red line: dust with frequency decorrelation (PySM model \texttt{d6}).}
    \end{center}
\end{figure}

The left panel of Fig.~\ref{fig-rbias} displays bands and polarization sensitivity of the various tested configurations. The ability to split the frequency band comes with a noise penalty caused by the smaller bands and by noise sub-optimality typical of bolometric interferometry (see section 5.3 of Mousset et al \cite{2020.QUBIC.PAPER2}). In the right panel we show the reconstructed tensor-to-scalar ratio as a function of the number of sub-bands, $n_\mathrm{sub}$, in two cases: (i) dust emission with a modified black body spectrum (green curve) and (ii) dust emission with frequency decorrelation (red curve)\footnote{We used PySM package to simulate the input sky. In particular the \texttt{d1} model for a modified black body dust emission and the \texttt{d6} model for dust with decorrelation.}. In the first case $r$ was set to 0.006, while in the second $r=0$. In both cases frequency decorrelation was not accounted for in component separation to mimic a situation in which the foregrounds contain unknown complexity.

The results show that if the dust spectrum is modeled appropriately then we recover the correct value of $r$, which does not depend on $n_\mathrm{sub}$. When the dust emission contains frequency decorrelation that is not considered in the component separation then we obtain a biased estimate that reduces by increasing $n_\mathrm{sub}$. This indicates that an imager integrating the signal in the bandwidth would not be able to distinguish a genuine detection from a biased one. With a bolometric interferometer, instead, it is possible to analyze the data by splitting the bands with various combinations of $n_\mathrm{sub}$ and detect the presence of a biased $r$.

\section{Experiment status and perspective}
\label{sec-status}

    A technological demonstrator (TD) of QUBIC has been installed at the end of 2022 at the observing site of Alto Chorrillo, in the Argentinean Andes at the altitude of about 5000\,m a.s.l. The QUBIC TD uses the same cryostat, cooling system, ﬁlters and general sub-system architecture as described above but with only 64 back-to-back horns and mirrors reduced according to the illumination of the 64 horns. It also uses a single 248 TES bolometer array operating at 150 GHz.

After commissioning activities, the QUBIC TD will demonstrate bolometric interferometry during 2024 by observing various sky patches with the aim to reconstruct the dust emission SED in total intensity and polarization. In parallel we will complete the development of the missing parts of the final instrument with a foreseen deployment during late 2024.

\section{Conclusions}
\label{sec-conclusions}

%

We presented QUBIC, an instrument designed to measure CMB $B$-modes using bolometric interferometry. This approach combines the sensitivity of TES detectors with systematic control typical of interferometers. A key feature is the ability to resolve sub-frequencies within the physical band, enabling spectral imaging not achievable by traditional CMB imagers with broader bandwidths. This capability is key to assess the robustness of a tensor-to-scalar ratio detection against foreground residual uncertainties. 



QUBIC technological demonstrator is operational at Alto Chorrillo in the Argentinean Andes at an altitude of approximately 5000\,m. It will observe the polarized sky at 150\,GHz for about a year in selected sky regions near the Galactic plane. The final instrument, equipped with two channels at 150 and 220\,GHz, is scheduled for deployment by late 2024. After three years of observations, it will constrain $r<0.03$ with 95\% C.L.

\bibliographystyle{woc}
\bibliography{biblio}

\end{document}